\documentclass[12pt]{article}
\usepackage{amsfonts, amscd}
\usepackage{amssymb}
\usepackage{eucal,eufrak}
\usepackage{verbatim}
\usepackage{latexsym}
\usepackage{epsfig}

\def\IR{{\mathbb R}}
\def\IC{{\mathbb C}}
\font\teneufm=eufm10
\font\seveneufm=eufm7
\font\fiveeufm=eufm5
\newfam\eufmfam
\textfont\eufmfam=\teneufm
\scriptfont\eufmfam=\seveneufm
\scriptscriptfont\eufmfam=\fiveeufm

\newcommand{\infinity}{\ensuremath{\infty}}
\newcommand{\dd}{\ensuremath{\partial}}

\newcommand{\fdot}{\partial_t f}

\newcommand{\fddot}{\partial^2_t f}
\newcommand{\drf}{\partial_r f}
\newcommand{\drrf}{\partial^2_r f}

\newcommand{\delr}{\triangle r}
\newcommand{\delt}{\triangle t}

\newcommand{\grad}{\raisebox{.5 ex}{\ensuremath{\bigtriangledown}}}

\begin{document}
\title{Slow Blow Up in the (2+1)-dimensional $S^2$ Sigma Model} 
\author{Jean Marie Linhart\\Applied Science Fiction\\8920 Business Park Drive
\\ Austin, TX 78759\\ jlinhart@asf.com}
\date{\today}
\maketitle

\begin{abstract}
We study singularity formation in spherically symmetric solitons of
the charge one sector of the (2+1) dimensional $S^2$ sigma model, also
known as $\IC P^1$ wave maps, in the adiabatic limit.  These equations
are non-integrable, and so studies are performed numerically on
radially symmetric solutions using an iterative finite differencing
scheme.  Analytic estimates are made by using an effective Lagrangian
cutoff outside a ball of fixed radius.  We show the geodesic
approximation is valid when the cutoff is applied, with the cutoff
approaching infinity linearly as the reciprocal of the initial
velocity.  Additionally a characterization of the shape of a time
slice $f(r,T)$ with $T$ fixed is provided.
\end{abstract}

\smallskip
\noindent Physics and Astronomy Classification: 02.30.Jr, 02.60.Cb
\medskip

\section{Introduction}
In this paper we study a hyperbolic partial differential equation that
develops a singularity in finite time.

The two-dimensional $S^2$ sigma model has been studied extensively
over the past few years in \cite{Leese}, \cite{PZ},
\cite{LPZ},\cite{Speight}, \cite{Ward}, \cite{Zakr}.  It is a good toy
model for studying two-dimensional analogues of elementary particles
in the framework of classical field theory.  Elementary particles are
described by classical extended solutions of this model,
called solitons.  This model is extended to (2+1) dimensions.  The
previous solitons are static or time-independent solutions, and then
the dynamics of these solitons are studied.  Since this model is not
integrable in (2+1) dimensions studies are performed numerically, and
analytic estimates are made by cutting off the model outside a radius
$R$.

The model can also be regarded as the continuum limit of an
array of Heisenberg ferromagnets.

The $S^2$ sigma model displays both slow blow up and fast blow up.  In
slow blow up, all relevant speeds go to zero as the singularity is
approached.  In fast blow up the relevant speeds do not go to zero as
the singularity is approached .  The charge 1 sector of the $S^2$
sigma model exhibits logarithmic slow blow up, whereas the charge 2
sector and the similar Yang Mills (4+1) dimensional model,
both investigated in \cite{Linhart} and \cite{Linhart3}, exhibit fast blow up.

The static Lagrangian density for the $S^2$ sigma model is given by
\[ L = \int|\grad\vec{\phi}|^2, \]
where $\vec{\phi}$ is a unit vector field.

In the dynamic version of this problem, where $\phi:\IR^{2+1}
\rightarrow S^2$, the Lagrangian is
\[ L = \int_{\IR^2} |\partial_t{\vec{\phi}}|^2 - |\grad\vec{\phi}|^2. \] 

Identifying $S^2 = \IC P^1 = \IC \cup \{\infinity\}$ we can rewrite
this in terms of a complex scalar field $u$:
\begin{equation} L = \int_{\IR^2} \frac{|\partial_t{u}|^2}{(1 + |u|^2)^2} - 
\frac{|\grad u|^2}{(1 + |u|^2)^2}.\label{cp1lag}\end{equation}

The calculus of variations on this Lagrangian in conjunction with
integration by parts yields the following equation of motion for the
$\IC P^1$ model:
\begin{equation}
(1 + |u|^2)(\dd_t^2u - \dd_x^2 u - \dd_y^2 u) = 2\bar{u}(|\dd_t u|^2 - |\dd_x u|^2
- |\dd_y u|^2) \label{genPDEb}\end{equation}
Here  $\bar{u}$ represents the complex conjugate of $u$.

The first thing to identify in this problem are the static solutions
determined by equation (\ref{genPDEb}).  These are outlined in
\cite{Ward} among others.  The entire space of static solutions can be
broken into finite dimensional manifolds $\mathcal{M}_n$ consisting of
the harmonic maps of degree $n$.  If $n$ is a positive integer, then
$\mathcal{M}_n$ consists of the set of all rational functions of $z =
x + iy$ of degree $n$.  For this chapter, we restrict our attention to
$\mathcal{M}_1$, the charge one sector, on which all static solutions have the
form
\begin{equation} u = \alpha + \beta(z + \gamma)^{-1}.\label{genc1soln}\end{equation}

In order to simplify, consider only solutions of the form
\[ \beta z^{-1}\]

The  geodesic approximation says that for slow
velocities solutions should evolve close to
\[ \frac{\beta(t)}{z}.\]
Instead of $\beta(t)$, we look for a real radially symmetric function $f(r,t)$, and find the evolution of: 
\[ \frac{f(r,t)}{z}.\]
The differences between the evolution of $f(0,t)$ and that predicted
in the geodesic approximation, and the deviation of $f(r,T)$ with $T$
fixed has from a horizontal line gives us a means to gauge how good
the geodesic approximation is.

It is straightforward to calculate the evolution equation for $f(r,t)$.  It is:

\begin{equation} \fddot = \drrf + \frac{3\drf}{r} - \frac{4r\drf}{f^2 + r^2} + 
\frac{2f}{f^2 + r^2}\left(\fdot^2 - \drf^2\right).\label{PDEb}\end{equation}

The static solutions for $f(r,t)$ are the horizontal lines $f(r,t) =
c$.  Here $c = \mbox{length\ scale}$. In the adiabatic limit motion
under small velocities should progress from line to line, i.e $f(r,t)
= c(t)$.  $f(r,t) = 0$ is a singularity of this system, where the
instantons are not well defined.  We use this to form a numerical
approximation to the adiabatic limit to observe progression from
$f(r,0) = c_0 > 0$ towards this singularity.

\section{Numerics for the $\IC P^1$ charge 1 sector model}

A finite difference method is used to compute the evolution of (\ref{PDEb})
numerically.  Centered differences are used consistently except for
\begin{equation} \drrf + \frac{3\drf}{r}.\label{instabpart}\end{equation}
In order to avoid serious instabilities in (\ref{PDEb}) this is
modeled in a special way.  Let
\[ \mathcal{L} f = r^{-3} \partial_r r^3 \partial_r f = \drrf + \frac{3\drf}{r}.\]
This operator has negative real
spectrum, hence it is stable.  The naive central differencing scheme
on (\ref{instabpart}) results in unbounded growth at the origin, but
the natural differencing scheme for this operator does not.  It is
\[ \mathcal{L}f \approx r^{-3}\left[ \frac{\left(r + \displaystyle{\frac{\delta}{2}}\right)^3\left(\displaystyle{\frac{f(r+\delta) - f(r)}{\delta}}\right) - \left(r-\displaystyle{\frac{\delta}{2}}\right)^3\left(\displaystyle{\frac{f(r) - f(r-\delta)}{\delta}}\right)}{\delta}\right].\]

Questions arise from \cite{LPZ} about the stability of the solitons
for this equation.  They found that a soliton would shrink without
perturbation from what should be the resting state.  This seems to be
a function of the numerical scheme used for those experiments. The
numerical scheme used here has no such stability problems.
Experiments show that a stationary solution is indeed stationary
unless perturbed by the addition of an initial velocity.

For a full analysis of the stability of this numerical scheme see
\cite{Linhart}.

With the differencing explained, we want to derive $f(r,t+\delt)$.  We
always have an initial guess at $f(r,t+\delt)$.  In the first time
step it is $f(r,t+\delt) = f(r,t) + v_0\delt$ with $v_0$ the initial
velocity given in the problem.  On subsequent time steps $f(r,t+\delt)
= 2f(r,t) - f(r,t-\delt)$.  This can be used to compute $\fdot(r,t)$
on the right hand side of (\ref{PDEb}).  Then solve for a new and
improved $f(r,t+\delt)$ in the differencing for the second derivative
$\fddot(r,t)$ and iterate this procedure several times to get
increasingly accurate values of $f(r,t)$.

There remains the question of boundary conditions.  At the origin
$f(r,t)$ is presumed to be an even function, and this gives
\[ f(0,t) = \frac{4}{3} f(\delr,t) - \frac{1}{3} f(2\delr,t).\]
At the $r = R$ boundary we presume that the function is horizontal
so $f(R,t) = f(R-\delr,t)$.

\section{Predictions of the Geodesic Approximation}

In \cite{PZ} the time evolution of the shrinking of solitons was
studied.  They arbitrarily cut off the Lagrangian outside of a ball of
radius $R$, to prevent logarithmic divergence of the integral for the
kinetic energy, and then  analyze what happens in the
$R\rightarrow \infinity$ limit.  In \cite{Speight} the problem of the
logarithmic divergence in the kinetic energy integral is solved by
investigating the model on the sphere $S^2$.  The radius of the sphere
determines a parameter for the size analogous to the parameter $R$ for
the size of the ball the Lagrangian is evaluated on in \cite{PZ}.

Our calculations here follow those done in \cite{PZ}. 

Equation (\ref{cp1lag}) gives us the Lagrangian for the general
version of this problem.  In the geodesic approximation or adiabatic
limit we have
\[ u = \frac{\beta}{z}\]
for our evolution.  If we restrict the Lagrangian to this space we get
an effective Lagrangian.  The integral of the spatial derivatives of
$u$ gives a constant, the Bogomol'nyi bound, and hence can be ignored.
If one integrates the kinetic term over the entire plane, one sees it
diverges logarithmically, so if $\beta$ is a function of time, the
soliton has infinite energy.

\begin{figure}[H]
\begin{center}
\epsfig{file=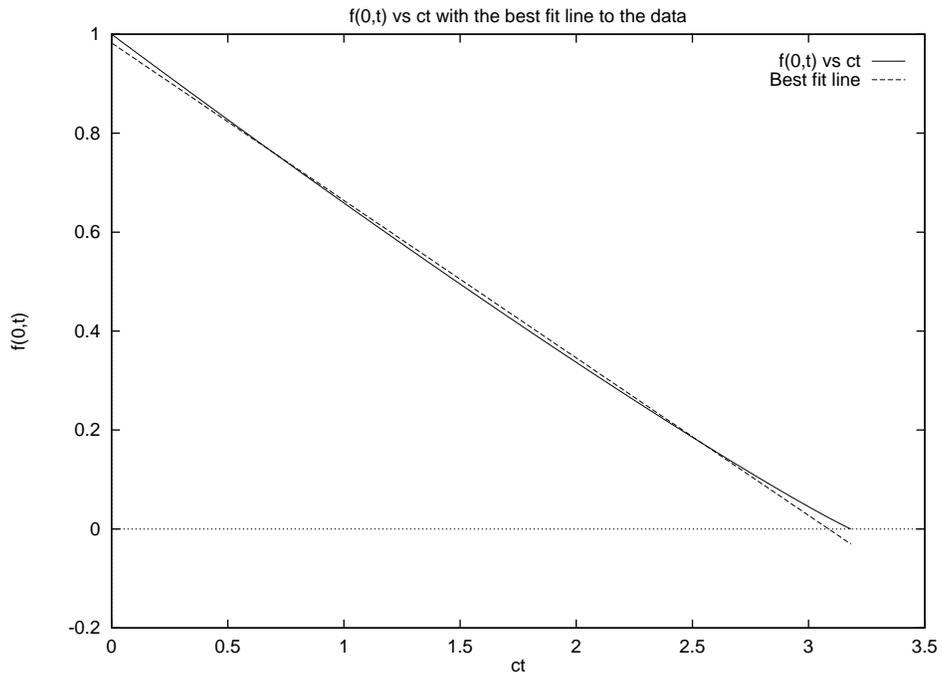}
\caption{Example plot of $f(0,t)$ vs $ct$ as predicted by the effective Lagrangian.}
\label{exampc1}
\end{center}
\end{figure}

Nonetheless, this is what we wish to investigate.  We cannot address
the entire plane in our numerical procedure either, hence we presume
that the evolution takes place in a ball around the origin of size
$R$.  If $\beta = f(r,t)$ shrinks to $0$ in time $T$, we need $R > T$.
Under these assumptions, up to a multiplicative constant, the
effective or cutoff Lagrangian becomes

\[ L = \int_0^R r dr \frac{r^2 \fdot^2}{(r^2 + f^2)^2} \]
which integrates to
\[ L = \frac{\fdot^2}{2} \left[ \ln\left(1 + \frac{R^2}{f^2}\right) -
\frac{R^2}{f^2 + R^2}\right]\] Since the potential energy is constant,
so is the purely kinetic Lagrangian, and
\[ \frac{\fdot^2}{2} \left[ \ln\left(1 + \frac{R^2}{f^2}\right) -
\frac{R^2}{f^2 + R^2}\right] = \frac{c^2}{2}, \]
with $c$ (and hence $c^2/2$) a constant.
Solving for $\fdot$ we obtain
\begin{equation} \fdot = \frac{c}{\sqrt{\left[\displaystyle{\ln\left(1 + 
\frac{R^2}{f^2}\right) - \frac{R^2}{f^2 + R^2}}\right]}}.
\label{fdoteqn} \end{equation}
Since we are starting at some value $f_0$ and evolving toward the singularity at $f = 0$ this gives:
\begin{equation} \int_{f_0}^{f(0,t)} d\:\! f 
\sqrt{\ln\left(1 + \frac{R^2}{f^2}\right) - \frac{R^2}{f^2 + R^2}} =
\int_0^t c dt.\label{origevo}\end{equation} The integral on the right
gives $ct$.  The integral on the left can be evaluated numerically
for given values of $R$, $f_0$ and $f(0,t)$.  A plot can then be
generated for $ct$ vs. $f(0,t)$.  What we really are concerned with is
$f(0,t)$ vs. $t$, but once the value of $c$ is determined this can be
easily obtained.  One such plot with $f_0 = 1.0$, $R = 100$ of
$f(0,t)$ vs $ct$ is given in Figure \ref{exampc1}.  This curve is
not quite linear, as seen by comparison with the
best fit line to this data which is also
plotted in Figure \ref{exampc1}.  The best fit line is obtained by a
least squares method.

\clearpage

\section{Evolution of the Origin Results}

The computer model was run under the condition that $f(r,0)
= f_0$ with various small velocities.  The initial velocity is
$\fdot(r,0) = v_0$, other input parameters are $R = r_{max}$,
$\delr$ and $\delt$.  

\smallskip
\noindent{\bf Evolution of $f(0,t)$}

\smallskip

The primary concern with the evolution of the horizontal line is the
way in which the singularity at $f(0,t) = 0$ is approached, because once again as $r \rightarrow \infinity$ equation (\ref{PDEb}) reduces to the linear
wave equation 
\[ \fddot = \drrf,\]
and so we expect the interesting behavior to occur near $r=0$.  The
model is run with initial conditions that $f(r,0) = f_0$, $\fdot(r,0)
= v_0$.  Other input parameters are $R= r_{\rm max}$, $\delr$ and
$\delt$.

\begin{figure}[H]
\begin{center}
\epsfig{file=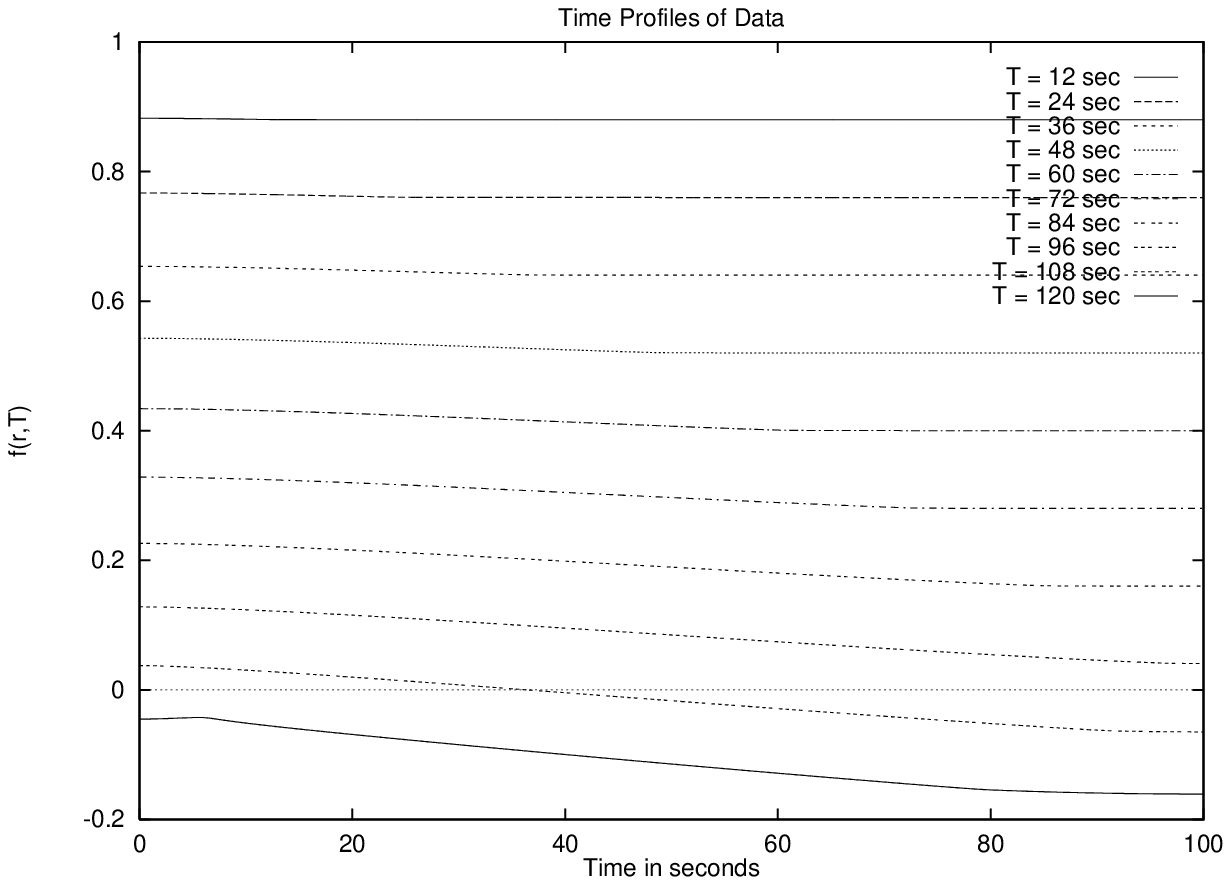}
\caption{Time slices $f(r,T)$}
\label{tscp1c1}
\end{center}
\end{figure}

The evolution of the initial horizontal line seems to remain largely
flat and horizontal, although there is some slope downward as time
increases.   This is shown in figure \ref{tscp1c1}.  

\begin{figure}[H]
\begin{center}
\epsfig{file=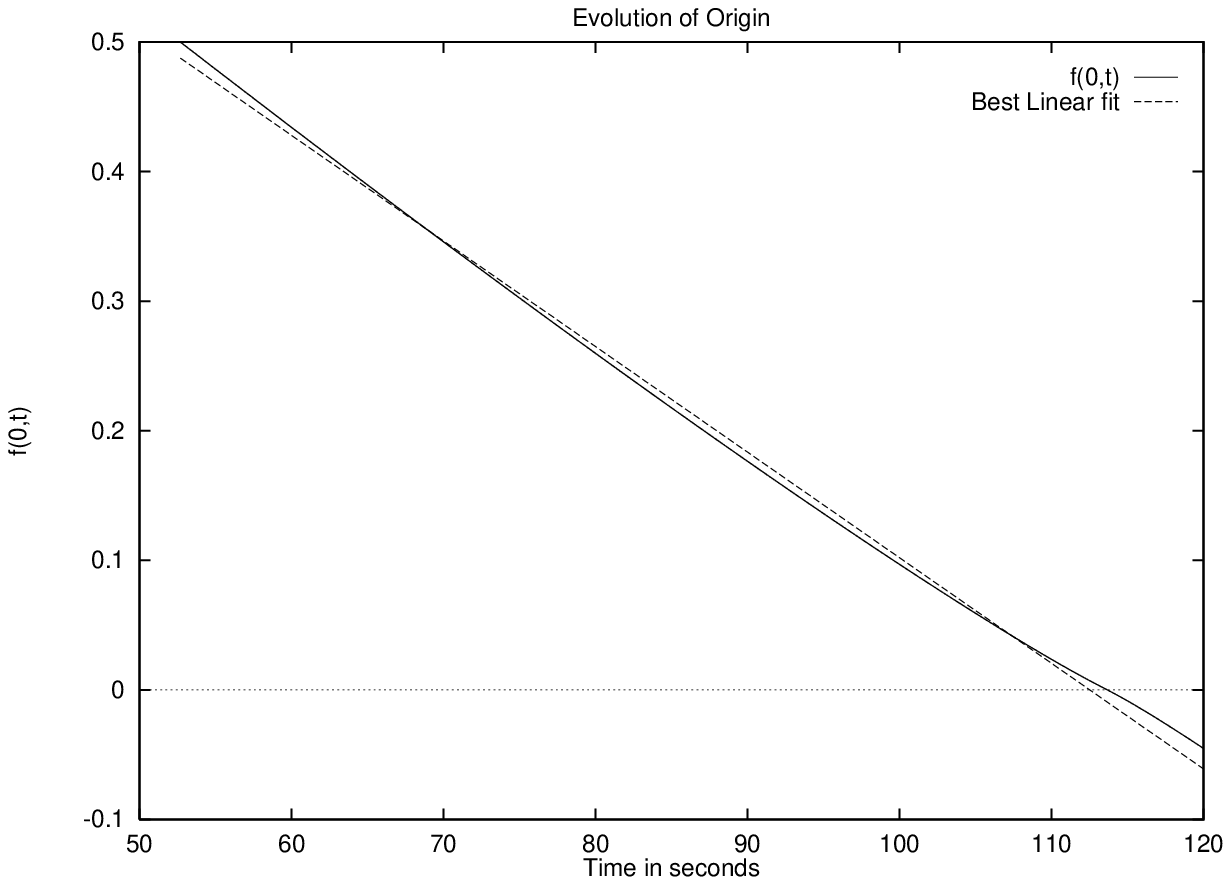}
\caption{Evolution of $f(0,t)$ is not quite linear.}
\label{ocp1c1}
\end{center}
\end{figure}

We track $f(0,t)$ as it heads toward this singularity, and
find that its trajectory is not quite linear, as seen in figure
\ref{ocp1c1}.  This is suggestive of the result obtained in the
predictions for this model.

We want to check the legitimacy of the result from
equation (\ref{origevo}) in chapter 3.2.  This requires a
determination of the parameters $R,$ the cutoff, and $c$.  We already
have $f_0$ and $f(0,t)$. To determine $R$ and $c$, observe from
equation (\ref{fdoteqn}) that:
\begin{equation} \frac{1}{\fdot^2} = \frac{\left[ 
\displaystyle{\ln\left(1 + \frac{R^2}{f^2}\right) - 
\frac{R^2}{f^2+R^2}}\right]}{c^2} \label{fdoteqnb}\end{equation}
Since $R$ is large and $f$ is small 
\[ \ln\left(1 + \frac{R^2}{f^2}\right) \approx \ln\left({\frac{R^2}{f^2}}\right)\]
and
\[ \frac{R^2}{f^2 + R^2} \approx 1.\]
Consequently we can rewrite equation (\ref{fdoteqnb}) as
\[ \frac{1}{\fdot^2} \approx \frac{\left[ 
\ln(R^2) - \ln(f^2) - 1\right]}{c^2}. \] 
The plot of
$\ln(f)= \ln(f(0,t))$ vs $1/{\fdot^2} = 1/{\fdot^2(0,t)}$
should be linear with the slope $m = 2/c^2$ and the intercept
$b = (2\ln(R) - 1)/c^2$.  Such a plot is easily obtained from the
model, and given slope and intercept, the parameters $c$ and $R$ are
easily obtained.

\begin{figure}[H]
\begin{center}
\epsfig{file=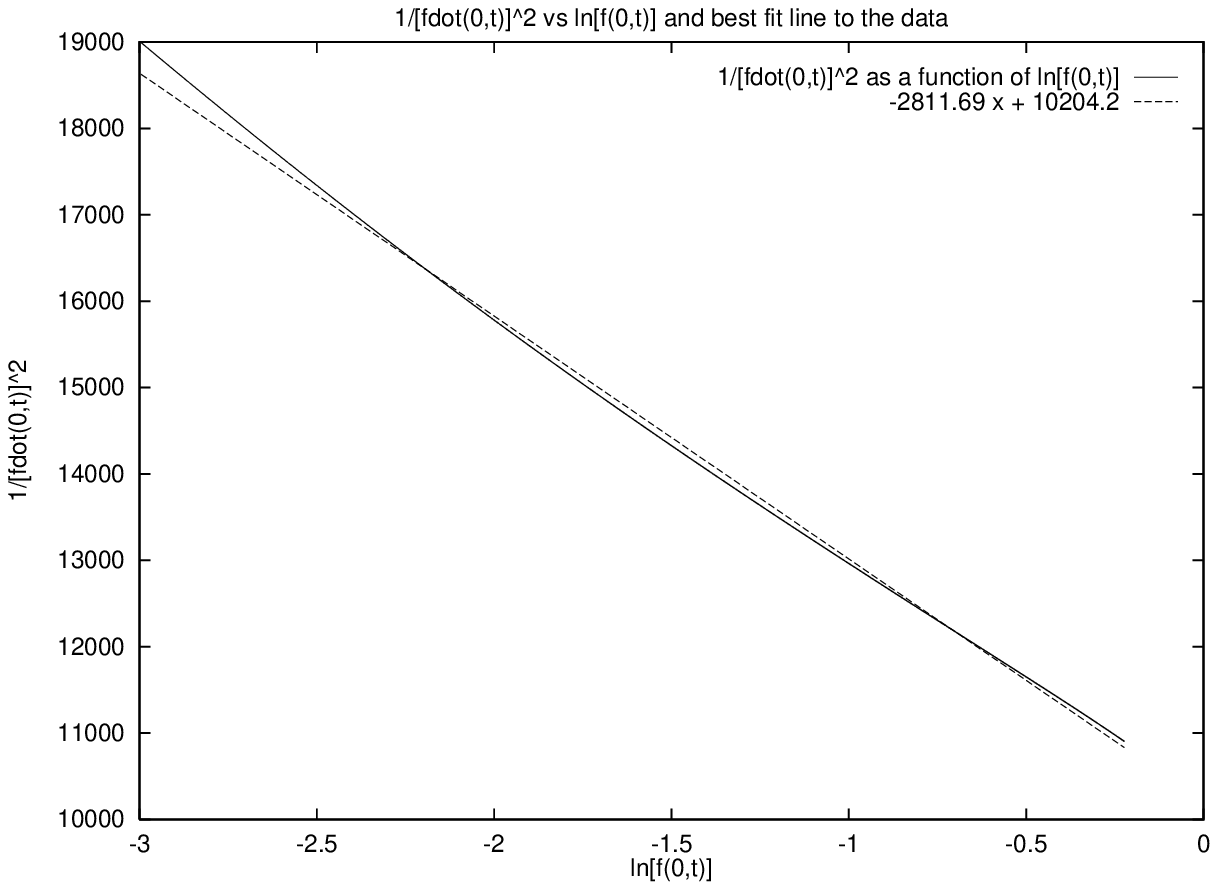}
\caption{Plot of $1/(\fdot)^2$ vs $\ln(f)$ and the best fit line to this data.}
\label{lnffdot}
\end{center}
\end{figure}

Figure \ref{lnffdot} is a plot of $\ln(f(0,t))$
vs. $1/\fdot^2(0,t)$, with initial conditions $\delr = 0.01$, $\delt
= 0.001$, $f_0 = 1.0$ and $v_0 = -0.01$.  It is easily seen that
although the plot of $\ln(f(0,t))$ vs. $1/\fdot^2(0,t)$ is nearly
straight, it is not quite a straight line.  This may indicate that the
values of $R$ and $c$ are changing with time.  

\begin{figure}[H]
\begin{center}
\epsfig{file=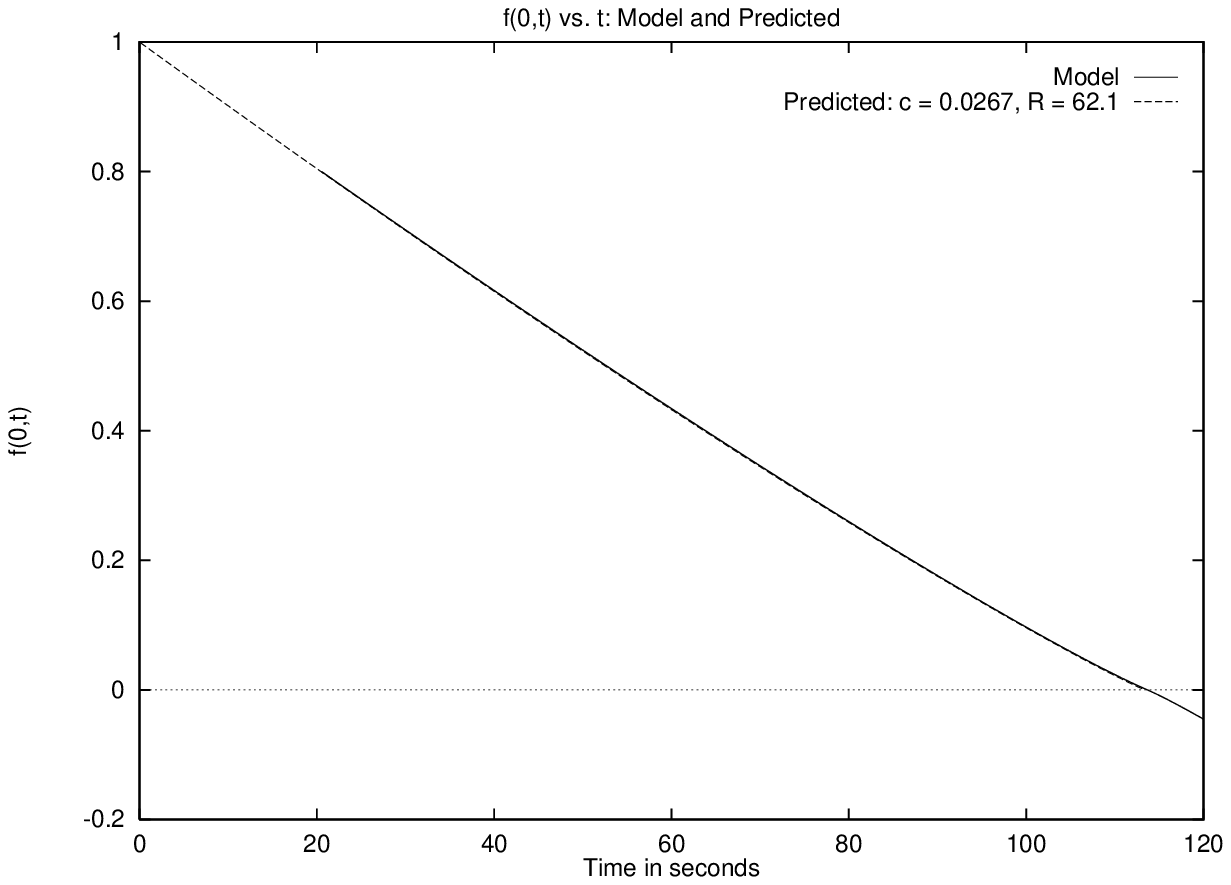}
\caption{Predicted course of $f(0,t)$ from equation (\ref{origevo}) and actual course of $f(0,t)$ vs. t.}
\label{fevol}
\end{center}
\end{figure}

The best fit line $y = mx + b$ has slope $m = -2810$ and $b = 10200$.
We have
\[c = \sqrt{\frac{-2}{m}}\]
and
\[R = \exp\left(-\frac{b}{m} + \frac{1}{2}\right).\]
This gives values $c = 0.0267$ and $R = 62.1$.  Using these values of
$c$ and $R$ in the calculation of equation (\ref{origevo}), we obtain the plot
of $f(0,t)$ vs $t$ given in Figure \ref{fevol}.  This is overlayed
with the model data for $f(0,t)$ vs. $t$ for comparison.  These two
are virtually identical.

This shows that the phenomenon of cutting off the Lagrangian outside
of a ball of radius $R$ is not just an artifact of necessity because
the full Lagrangian is divergent, but an inherent feature of this
system.

\begin{figure}[H]
\begin{center}
\epsfig{file=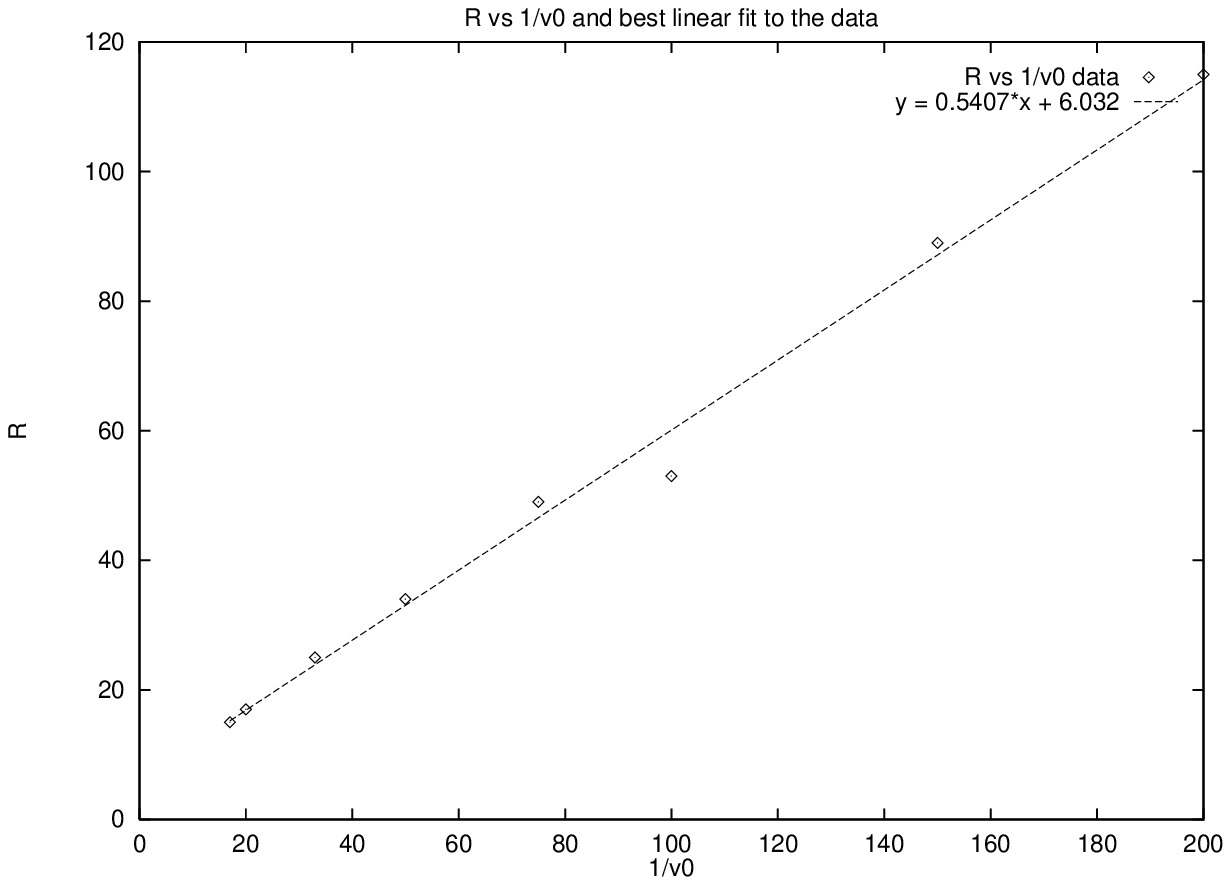}
\caption{ R vs 1/v0 and the best fit line.}
\label{crvfit}
\end{center}
\end{figure}

Table \ref{crvtab} contains the data for $c$ and $R$ vs. change in
the initial velocity $v_0$, under the initial conditions $f_0 = 1.0$,
$\delr = 0.01$ and $\delt = 0.001$.  The data for $R$ varying with $1/v_0$ 
fits well to the line
\[ y= 0.5407 x + 6.032.\]
This fit is shown in figure {\ref{crvfit}}.  A linear fit makes sense, since
as the velocity tends toward zero, we expect the cutoff to head toward
infinity.  

Table \ref{crftab} containing the data for $c$ and $R$ vs. change in
$f_0$, the initial height.  The parameter $R$ varies close to linearly
with $f_0$, while the parameter $c$ remains nearly constant, changing
by less than $7\%$ over the course of the runs.

\begin{table}[H]
\begin{center}
\caption{$c$ and $R$ vs. $v_0$.
$f_0 = 1.0$.}
\label{crvtab}
\[ \begin{array}{{r}{l}{r}} v_0  & c & R \\
-0.005&   0.0145& 115\\
-0.00667& 0.0187&  89\\
-0.01&    0.0263&  53\\
-0.0133&  0.0342&  49\\
-0.02&    0.0485&  34\\
-0.03&    0.0683&  25\\
-0.05&    0.104&   17\\
-0.06&    0.121&   15 \\
\end{array}\]
\end{center}
\end{table}

\begin{table}[H]
\begin{center}
\caption{$c$ and $R$ vs. $f_0$.
$v_0 = -0.01$.}
\label{crftab}
\[ \begin{array}{{r}{l}{r}} f_0  & c & R \\
1.0&   0.0267&    62\\
2.0&   0.0263&   108\\
3.0&   0.0260&   150\\
4.0&   0.0259&   190\\
\end{array}\]
\end{center}
\end{table}

\clearpage

{\bf Characterization of time slices $f(r,T)$}

\begin{figure}[H]
\begin{center}
\epsfig{file=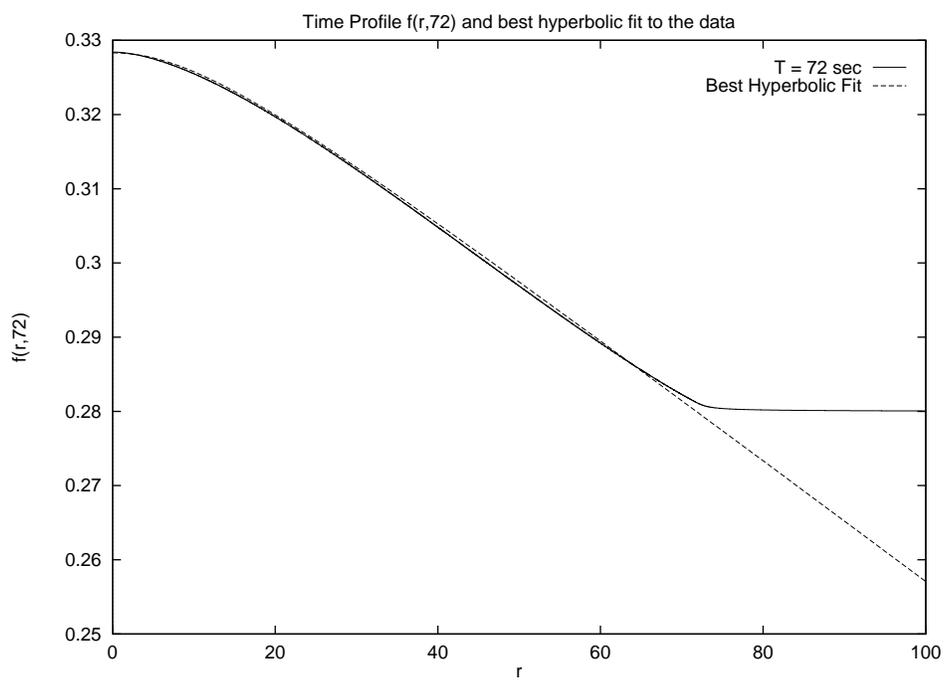}
\caption{Time slices evolve hyperbolic bump at origin.}
\label{hcp1c1}
\end{center}
\end{figure}

Making a closer inspection of the time profiles $f(r, T)$ with $T$
fixed as in Figure \ref{tscp1c1}, one may observe that the initial part of
the data is close to a hyperbola as seen in Figure \ref{hcp1c1}.
The best hyperbolic fit is determined by a least squares method.

The equation for the hyperbola is 

\[ \frac{(y-k)^2}{b^2} - \frac{x^2}{a^2} = 1.\]

One would naturally ask about the evolution of the hyperbolic
parameters $a$ and $b$ with time, however, neither of these is
particularly edifying.  A simple calculation shows that $k$ should
follow $f(0,t)$ closely if $b$ is small, as it is.  

\begin{figure}[H]
\begin{center}
\epsfig{file=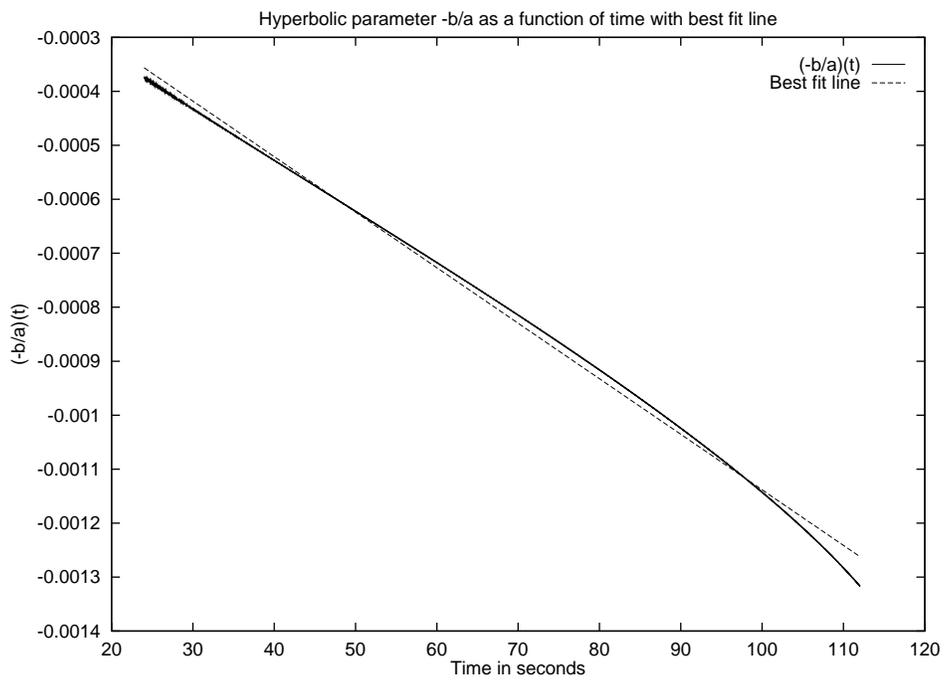}
\caption{Plot of hyperbolic parameter $-b/a$ vs time, $f_0 = 1.0$, $v_0 = -0.01$.}
\label{hba}
\end{center}
\end{figure}

The evolution of $-b/a$ gives the slope of the asymptotic line to the
hyperbola, and this evolution is close to linear as seen in Figure
\ref{hba}.  

\clearpage

\section{Conclusions}

In \cite{LPZ}, solitons are found to be numerically unstable.  A
solution of the form
\[ \frac{\beta}{z} \]
shrinks spontaneously under their numerical procedure.  This does not
occur in our numerical implementation of the $S^2$ sigma model.  The
static solutions do not evolve in time unless given an initial rate of
shrinking.  Further, stability and convergence analysis of two of the
numerical procedures is provided in \cite{Linhart}.

We use a method analogous to that in \cite{PZ} of cutting off the
Lagrangian outside of a ball of radius $R$, and we find an explicit
integral for the shrinking of the soliton, dependent on two
parameters: $c$ which is a function of the kinetic energy, and $R$
which is the size of the ball on which we evaluate the Lagrangian.
Once these are specified, this integral gives the theoretical
trajectory of the soliton.  We find an explicit integral for the
shrinking of a soliton where the Lagrangian is cut off outside of the
ball of radius $R$, and when $R$ is calculated (along with the
parameter $c$), the shrinking seen in the numerical model matches that
predicted.  The dependence of $R$ on the initial conditions appears to
be linear with the initial velocity.  The validity of the geodesic
approximation with a cutoff $R$ is shown by this, and the cutoff is an
essential part of this system.

In addition to this, the shape of a time slice $f(r,T)$, with $T$
fixed, is characterized by hyperbolic bumps at the origin.

\section{Acknowledgments}

I would like to thank my dissertation supervisor, Lorenzo Sadun, for
his constructive comments and suggesstions for additional lines of
research and improvement on this manuscript and my dissertation.

\bibliography{art}
\end{document}